\documentclass[preprint,floats,aps,showpacs]{revtex4}

\usepackage{graphicx,amsmath,amssymb}

\newcommand{\<}{\langle}
\renewcommand{\>}{\rangle}
\newcommand{\be}{\begin{equation}}
\newcommand{\ee}{\end{equation}}
\newcommand{\bea}{\begin{eqnarray}}
\newcommand{\eea}{\end{eqnarray}}

\def\ox{\overline{x}}
\def\oy{\overline{y}}
\def\Qh{\widehat{Q}}
\def\Mh{\hat{M}}
\def\rh{\widehat{\rho}}
\def\eh{\hat{\epsilon}}

\begin{document}

\title{On the cooling-schedule dependence of the dynamics\\
of mean-field glasses}

\author{Andrea Montanari}

\affiliation{Laboratoire de Physique Th\'{e}orique de l'Ecole Normale
Sup\'{e}rieure\footnote {UMR 8549, Unit{\'e}   Mixte de Recherche du
Centre National de la Recherche Scientifique et de l' Ecole Normale
Sup{\'e}rieure. }, 24, rue Lhomond, 75231 Paris CEDEX 05, France}

\author{Federico Ricci-Tersenghi}

\affiliation{Dipartimento di Fisica, INFM (UdR Roma I and SMC center)\\
Universit\`{a} di Roma ``La Sapienza'', P. Aldo Moro 2, I-00185 Roma,
Italy}

\begin{abstract}
The low temperature phase of discontinuous mean-field spin glasses is
characterized by the appearance of an exponential number of metastable
states. Which ones among these states dominate the out-of-equilibrium
dynamics of these systems?

In order to answer this question, we compare high-precision numerical 
simulations of a diluted $p$-spin model with a cavity computation of 
the threshold energy.  Our main conclusion is that the aging dynamics 
is dominated by different layers of metastable states depending on the 
cooling schedule.  In order to perform our analysis, we present a method 
for computing the marginality condition of diluted spin glasses at 
non-zero temperature.
\end{abstract}


\maketitle

\section{Introduction and results}
\label{Introduction}

Upon cooling, glass-forming liquids show a dramatic slowing-down of
their relaxational dynamics.  In mean-field models, this phenomenon is
caricatured by a {\it weak ergodicity breaking} phase transition (the
so called ``dynamic phase transition'') which occurs at some
critical temperature $T_{\rm d}$~\cite{DynamicsReview,KirkpatrickThirumalai1}.
Surprisingly
enough, this phenomenon could be relevant in understanding the
behavior of local search algorithm for hard computational
problems~\cite{CSReview,PTAC}.  The time-complexity of these
algorithms suddenly explodes when some macroscopic parameter
describing the instances crosses a critical value.

Despite their apparent simplicity, the dynamics of mean-field models
is still poorly understood in many aspects.  Consider, for instance,
the following couple of {\it Gedanken} experiments.  In the first one,
the system is at infinite temperature at time $t=0$: $T(0) = \infty$,
and is suddenly quenched below the dynamic transition: $T(t) =
T<T_{\rm d}$ for $t>0$. At some time $t_{\rm f}$ we measure the observable
${\cal O}$ and the experiment is finished. We are interested in the
asymptotic behavior:
\begin{eqnarray}
\<{\cal O}\>_{\rm quench} \equiv \lim_{t_{\rm f}\to\infty}
\<{\cal O}(t_{\rm f})\>_{T}\, . \label{QuenchDefinition}
\end{eqnarray}
In the second experiment we choose a cooling schedule, i.e.\ a smooth
function $T_{\rm sch}(\tau)$, $0\le \tau\le 1$, with $T_{\rm
sch}(0)=\infty$ and $T_{\rm sch}(1) = T$. We start from $T(0) =
\infty$ and slowly cool down the system, keeping it at temperature
$T(t) = T_{\rm sch}(t/t_{\rm f})$ for $0\le t\le t_{\rm f}$.  Then we
measure the same observable ${\cal O}$ and consider the limit
\begin{eqnarray}
\<{\cal O}\>_{\rm cool} \equiv \lim_{t_{\rm f}\to\infty}
\<{\cal O}(t_{\rm f})\>_{T_{\rm sch}(t/t_{\rm f})}\,.
\label{CoolDefinition}
\end{eqnarray}
In this paper we address the question:
\begin{eqnarray}
\<{\cal O}\>_{\rm quench} \stackrel{?}{=} \<{\cal O}\>_{\rm cool}\,.
\label{Question}
\end{eqnarray}
Notice that both in Eqs.~(\ref{QuenchDefinition}) and
(\ref{CoolDefinition}), the thermodynamic limit is assumed to be taken
{\it before} the long-time limit. In the opposite case the answer
would be trivially positive.

It turns out that the two quantities are in general different:
$\<{\cal O}\>_{\rm quench} \neq \<{\cal O}\>_{\rm cool}$.  
In the following we shall focus on the internal energy,
${\cal O} = {\cal H}(\sigma)$ and we will show that in this case
$\<{\cal H}(\sigma)\>_{\rm quench} > \<{\cal H}(\sigma)\>_{\rm cool}$
strictly. This
result is quite surprising. The bulk of analytical results in this
field comes, in fact, from the solution of the fully connected
$p$-spin spherical model~\cite{CrisantiSommers_Statics,
CrisantiSommers_Dynamics,CugliandoloKurchanPspin}.  
In this case $\<{\cal O}\>_{\rm quench} = \<{\cal O}\>_{\rm cool}$.

Like other properties of the out-of-equilibrium dynamics of
mean-field models, the result $\<{\cal O}\>_{\rm quench} \neq 
\<{\cal O}\>_{\rm cool}$ can be interpreted in connection with the structure 
of metastable states. It is worth recalling this connection here. 
Below $T_{\rm d}$ the Boltzmann measure splits into
an exponential number of metastable states. 
Formally~\footnote{For the specialist: Throughout this paper we identify
`states' by the condition that two typical configuration in the same state
have overlap greater or equal than $q_{\rm min}$. Two configurations in different states have instead zero overlap. Here $q_{\rm min}$
is the smallest non-zero overlap in the support of the Parisi order parameter
$P(q)$ (notice that such a definition is meaningful only for discontinuous 
models).
Sometimes other names (`families', `clusters', etc) are used 
for the same objects to emphasize that they can be decomposed further.}
\begin{eqnarray}
\<{\cal O}\>_{\rm eq} \approx \sum_{\alpha}w_{\alpha} \<{\cal O}\>_{\alpha}\, .
\label{Decomposition}
\end{eqnarray}
Such a splitting has been proven rigorously for some completely
connected $p$-spin models \cite{Talagrand}.
Two different states have different values of the extensive 
observables (energy, magnetization, etc). One can therefore 
construct constrained Boltzmann measures using only those states 
which have a definite value of one of these observables. For instance one
can constrain on the energy
\begin{eqnarray}
\<{\cal O}\>_{e} \propto  \!\!\!\!\!
\sum_{\alpha\; :\;\<{\cal H}\>_{\alpha} \approx Ne}\!\!\!\!\!\!
w_{\alpha}\;\; \<{\cal O}\>_{\alpha}\, .\label{Constrained}
\end{eqnarray}
It is a simple property of mean field models that purely relaxation
dynamics does not necessarily converge to (Boltzmann) equilibrium.
Consider for instance a Curie Weiss model at low temperature in
a small positive magnetic field. If the dynamics is initiated from a 
(sufficiently) negative magnetization configuration, it will converge to 
a negative magnetization state unless one waits a time exponential in the 
size of the system~\cite{Langer}.

Unlike in the Curie Weiss model, in glassy systems it is difficult
to select a particular metastable state by switching on a field in the
appropriate direction. In general, the appropriate out-of-equilibrium
measure will rather resemble a mixture of metastable states.
A fascinating hypothesis on mean-field spin glasses 
is that relaxation dynamics converges indeed
(as long as one-time observables are concerned)
to a constrained measure of the form (\ref{Constrained}). If one 
accepts this hypothesis, the open question is: which of the many possible
constrained measures does the dynamics converge to? Our results
imply that the answer depends upon the path along which the system is 
driven into its low-temperature phase. Viceversa, this path can be
tailored in such a way to select different constrained measures.

\begin{figure}
\begin{tabular}{cc}
\includegraphics[width=0.45\textwidth]{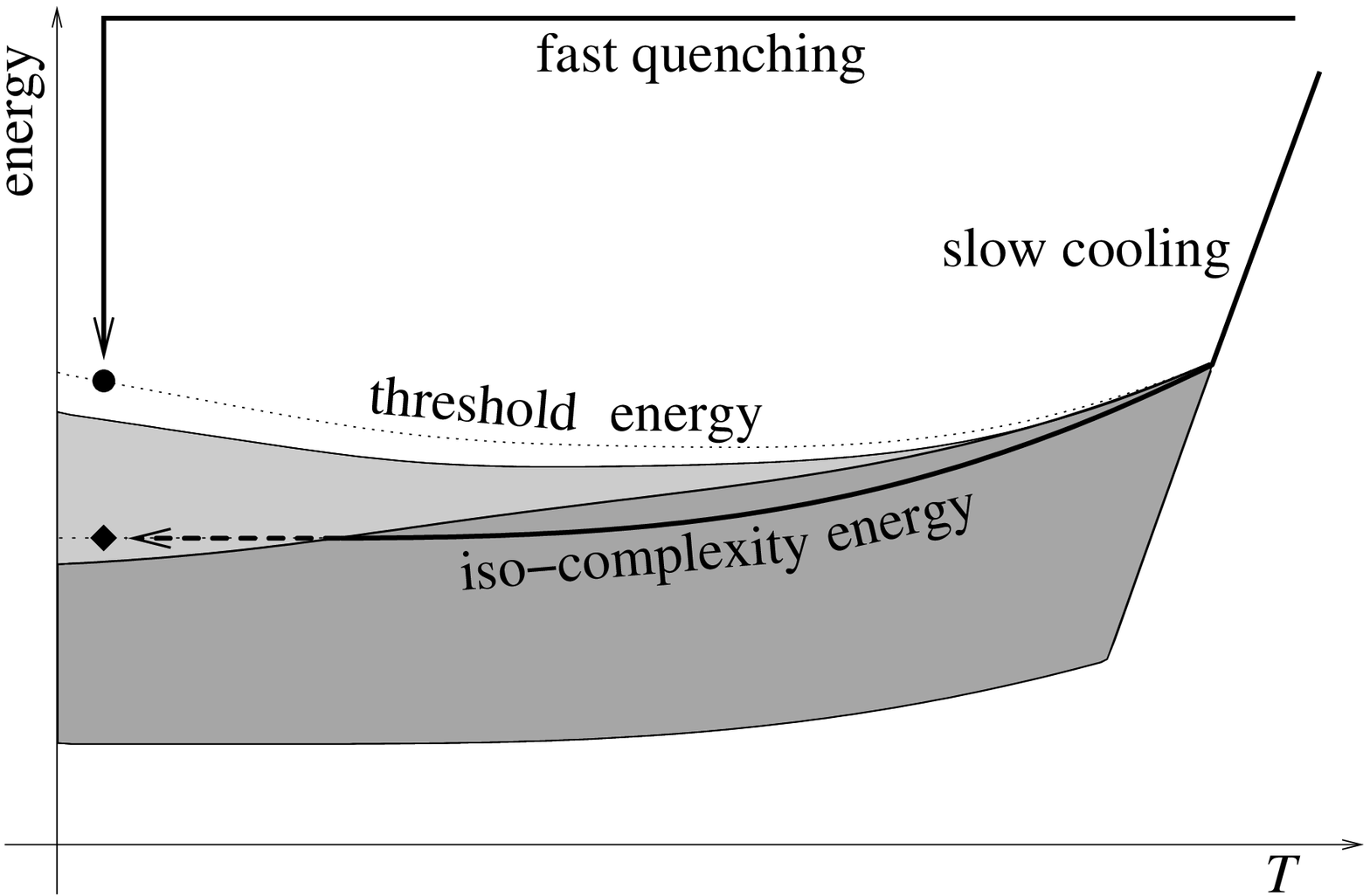}&\hspace{1.cm} 
\includegraphics[width=0.45\textwidth]{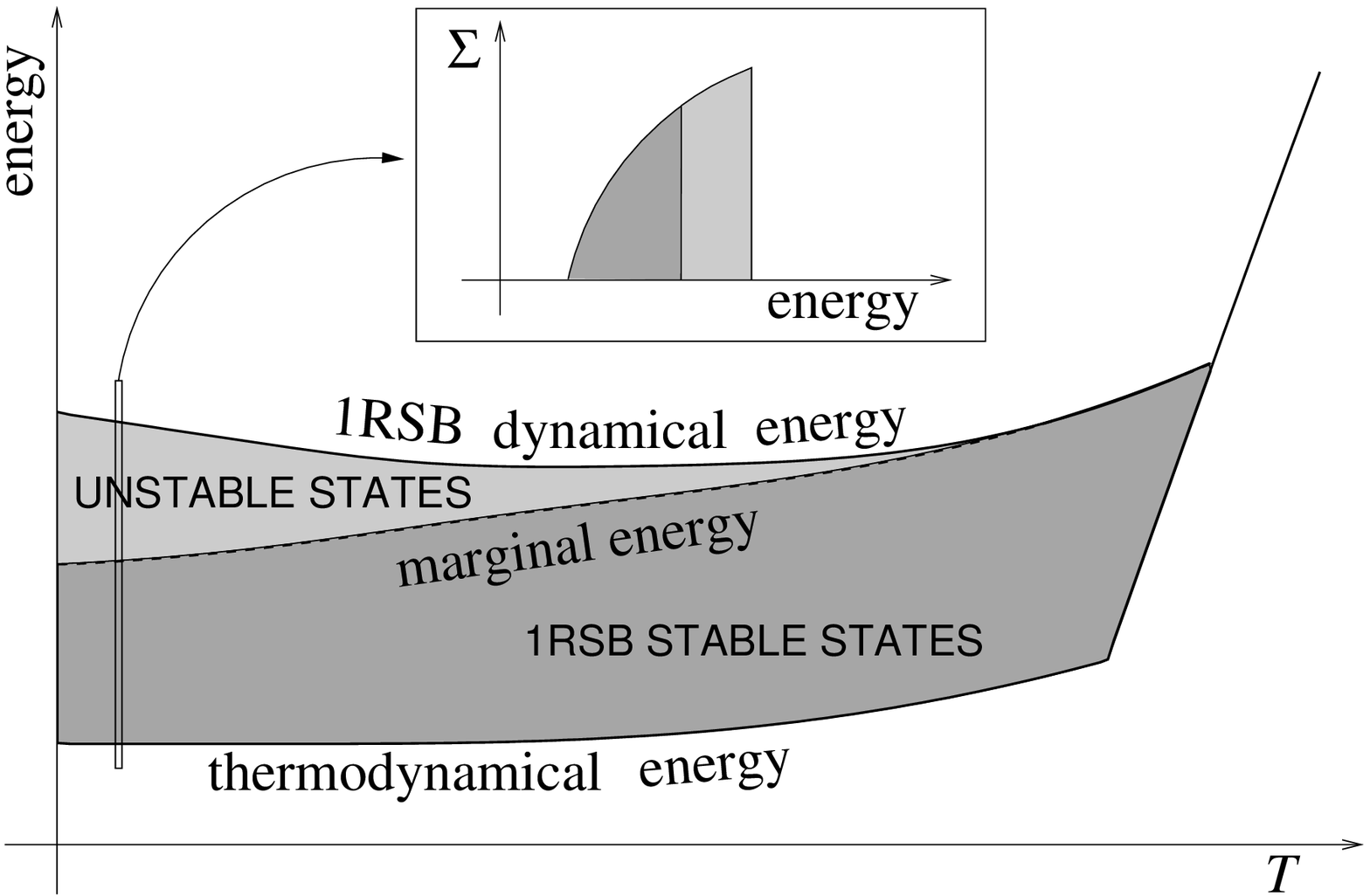}
\end{tabular}
\caption{A pictorial view of the temperature evolution of metastable
states in a discontinuous mean-field model, {\it without} temperature 
chaos. The gray areas correspond to those regions in the 
temperature-energy plane, such that an exponential number of metastable states
is present, within a 1RSB calculation.  
Dark gray corresponds to states which are stable
with respect to FRSB, while light gray corresponds to unstable
states.
On the left we depict the trajectory of a quench and a cooling
protocols. On the right (inset) we plot the 1RSB complexity (i.e.
the entropy of metastable states) at a particular temperature.
By `threshold states', we mean here the highest energy states such that 
the complexity is strictly positive. In general FRSB must be used
to describe these states, as well as to calculate the iso-complexity energy
in the  light gray region (dashed line).}
\label{Pictorial}
\end{figure}
Figure \ref{Pictorial} sketches the mechanism for selecting metastable
states at different energy densities by changing the cooling schedule.
The evolution of the energy density and its asymptotic value
are traced. In the inset we show the complexity (i.e. the logarithm of
the number of metastable states) at the final temperature. It turns 
out that cooling slowly the system through the dynamic transition 
one is able to avoid the higher energy (and most numerous) metastable states.
This is not the case after a rapid quench. We also show the 
marginal energy density which correspond to the appearance
of a full hierarchy of `nested' decompositions of the form 
(\ref{Decomposition}). In spin glass jargon, this corresponds to full 
replica symmetry breaking (FRSB).
Let us notice that these sketches, as well as the examples in the following,
refer to cases in which a thermodynamic FRSB transition does not exist
(in other words the marginal energy line does not cross the thermodynamical 
energy line). This is not, however, a fundamental  restriction of the theory.

Already Crisanti, Horner and Sommers~\cite{CrisantiSommers_Dynamics}
noticed that the asymptotic value of one-time observables depends upon
the cooling schedule for the spherical $p$-spin model in {\it non-zero
magnetic field}. This phenomenon was further investigated  in a 
non-homogeneous spherical model by Barrat, Franz and Parisi
\cite{FranzParisi,BarratFranzParisi}.
The crucial point is that the $p$-spin spherical model is an exceptional 
case as far as the structure of high-lying metastable states is considered.

Although several conclusions of \cite{BarratFranzParisi} agree with our 
findings, they relied on some uncontrolled assumptions and interpretations.
The generality of their results wasn't fully appreciated. 
In this paper we revisit the same problem in a context~\cite{NostroInst}, 
where numerical simulations can be carefully compared with analytical
results, and assumptions can be checked. 
Moreover, the model considered is a close (although simpler) relative
of a large family of hard optimization problems.

Our conclusions are based on an extensive numerical study of a diluted
Ising spin glass with $p$-spin interactions.  The dynamics of these
models can be simulated with a smaller computational effort with
respect to their fully-connected counterparts. Moreover, the analysis
of their static properties has being rapidly developing in the last
few years.  This paper contains one further step in this
direction. Developing the ideas of Ref. \cite{NostroInst}, we show how
to compute the marginality condition of one-step replica-symmetry
breaking (1RSB) solutions for diluted models at non-zero temperature.

The paper is organized as follows. In Section \ref{ModelSection} we
define the family of models to be investigated and give some
analytical results concerning the metastable states of these
models. Moreover we explain how the marginality condition is computed.
In Sec.~\ref{NumericalSection} we present our numerical simulations
and compare them with the static calculations mentioned above.
A few comments on these results are presented in
Sec.~\ref{DiscussionSection}. 
In Appendix \ref{ExplicitFormulae} we collect some explicit formulae for
the model of Sec.~\ref{ModelSection}.  Finally in the
App.~\ref{LargeConnectivity}, we show that, in the large connectivity
limit, our computation of the marginality condition coincides with the
usual replicon result \cite{Gardner}.
%
%
\section{The model and its marginality condition}
\label{ModelSection}

We shall consider the family of diluted spin glasses defined by the
Hamiltonian
\begin{eqnarray}
{\cal H}(\sigma) = -\sum_{(i_1 \dots i_p)\in {\cal G}} J_{i_1 \dots
i_p} \sigma_{i_1} \dots \sigma_{i_p}\; .\label{Hamiltonian}
\end{eqnarray}
where ${\cal G}$ is the hypergraph of interactions, i.e.\ a set of $M$
among the $\binom{N}{p}$ possible $p$-uples (with $p\ge 3$) of the $N$
spins. We shall consider hypergraphs ${\cal G}$ with fixed connectivity: each
spin is supposed to participate to $c=l\!+\!1$ interaction terms
(hereafter $c \ge 3$).  The graph ${\cal G}$ is drawn randomly, with 
uniform probability, among all such hypergraphs (i.e. among all the
hypergraphs on $N$ vertices with connectivity $c$, and $p$-vertices edges).
Finally we shall assume the couplings
$J_{i_1 \dots i_p}$ to take the values $\pm 1$ with equal probability.

This model was already studied in
Refs.~\cite{FranzEtAlFerromagn,FranzEtAlExact} within a 1RSB
approximation. As the temperature is lowered it undergoes a dynamic
phase transition at $T_{\rm d}>0$ and, if $c>p$, a static phase
transition at $T_{\rm s}$, with $0<T_{\rm s}<T_{\rm d}$.  For $l\ge
l_{\rm G}(p)$ a second static phase transition takes place at some
temperature $T_{\rm G}$: the $T<T_{\rm G}$ phase is characterized by
full replica-symmetry breaking. In Ref. \cite{NostroInst} we
studied the $T=0$ limit and found $l_{\rm G}(3) = 10$.  In
Tab.~\ref{PhaseTransitionTable} we report the values of $T_{\rm s}$
and $T_{\rm d}$ for $p=3$ and a few values of $l<10$.

\begin{table}
\centering\begin{tabular}{|c|c|c|c|c||c|c||c|}
\hline
\ $c$\ \ & $T_{\rm s}$ & $T_{\rm d}$ & $\Sigma_{\rm d}(T_{\rm d})$ &
 $e_{\rm d}(T_{\rm d})$ & $T_{\rm max}$ & $\Delta T$ & $T_{\rm quench}$ \\
\hline
\hline
 4 & 0.655(5) & 0.757(5) & 0.0540(10) & -1.157(2) & 1.0 & 0.01 & 0.4, 0.5, 0.6 \\
 5 & 0.849(5) & 0.936(5) & 0.0445(10) & -1.314(2) & 1.5 & 0.01 & 0.6, 0.7, 0.8 \\
 8 & 1.25(1)  & 1.345(5) & 0.0335(10) & -1.683(5) & 2.0 & 0.02 & 0.8, 1.0, 1.2 \\
\hline
\hline
\end{tabular}
\caption{Critical temperatures for the static ($T_{\rm s}$) and
dynamic ($T_{\rm d}$) transitions of the model (\ref{Hamiltonian})
as obtained from a population dynamics solution of the 1RSB
equations. We also report the complexity ($\Sigma_{\rm d}$) and energy
($e_{\rm d}$) of threshold states at the dynamic transition. Quoted
errors include both statistical and systematic uncertainties. The two
central columns refer to parameter values used in the numerical
cooling experiments, while the last column contains the temperatures
used for the quenching experiments.}
\label{PhaseTransitionTable}
\end{table}
At a fixed temperature $T=1/\beta<T_{\rm d}$, the Gibbs measure
decomposes over an exponential number of pure (metastable) states. The
number of states at a given free energy density $f$ is given, at the
leading exponential order, by ${\cal N}_\beta(f) \sim
\exp\{N\Sigma_{\beta}(f)\}$. In 1RSB approximation, the complexity
$\Sigma_\beta(f)$ is obtained \cite{MonassonMarginal} as the Legendre
transform of the $m$ replicas free energy $m\, \phi(m,\beta)$ with
respect to the parameter $m$:
\begin{eqnarray}
\Sigma_{\beta}(f) =\left. m\beta\, f - m\, \phi(m,\beta)\right|_{\beta
f = \partial_m\,[m\phi]}
\end{eqnarray}
The parameter $m$ can be varied in the range $m_{\rm d}(T)<m<m_{\rm
s}(T)$.  This corresponds to selecting states of free-energy densities
$f_{\rm d}(T) > f>f_{\rm s}(T)$. Any other observable can be computed
over metastable states of a given free energy, by properly tuning $m$.
The internal energy, for instance, decreases from $e_{\rm d}(T)$ to
$e_{\rm s}(T)$, as $f$ varies from $f_{\rm d}(T)$ to $f_{\rm s}(T)$.
In Tab.~\ref{CurvesTable} we report the curves $e_{\rm s}(T)$ and
$e_{\rm d}(T)$ for $p=3$ and a few values of $l$, see also 
Figs.~\ref{cool_all}, \ref{cool_c5_zoom}.  
The 1RSB saddle point equations were solved using the
population dynamics algorithm of Ref. \cite{MezardParisiBethe}.

As shown in Ref. \cite{NostroInst}, high-energy metastable states are,
quite generically, unstable towards FRSB. Here we want to compute the
stability threshold $m_{\rm G}(T)$ and the corresponding internal
energy $e_{\rm G}(T)$ at finite temperature $T$.  We proceed by
determining the instability toward two-steps replica symmetry breaking
(2RSB), which in turn is expected to imply the instability towards
FRSB. We shall present the method in a general setting and give some
numerical results, cf. Tab.~\ref{CurvesTable}. We refer to App.~\ref{ExplicitFormulae} for explicit formulae in the case of the
Hamiltonian (\ref{Hamiltonian}).

Let ${\cal S}$ be the space of normalized measures $\rho(x)$ over the
real numbers. The 2RSB order parameter for the model
(\ref{Hamiltonian}) is a probability distribution $Q[\rho]$ over
${\cal S}$.  This is slightly simpler than the most general 2RSB order
parameter for a diluted model because we assumed all the sites to be
equivalent~\cite{FranzEtAlExact}. Notice in fact the finite
neighborhood of any two sites $i$ and $j$ are identical (up to a gauge
transformation $\sigma_i\to\tau_i\sigma_i$) in the thermodynamic
limit.

We consider now a generic model such that, once an interaction term has 
been taken away, each adjacent spin interacts with
$k$ other ones [to make contact with the Hamiltonian
(\ref{Hamiltonian}) one should take $k=l(p-1)$].  The 2RSB saddle
point equations for such a model have the form
\begin{eqnarray}
Q[\rho] = \frac{1}{\cal Z}\int\!\prod_{i=1}^{k}\! dQ[\rho_i] \,\,
z[\rho_1,\dots,\rho_k]^{m_1/m_2}\,\,
\delta\left[\rho-\rho_0[\rho_1,\dots,\rho_k]\right]\, ,
\label{2RSBSaddlePoint}
\end{eqnarray}
where $m_1<m_2$ are the two breaking parameters required for 2RSB
\cite{SpinGlass}.  The functional $\rho_0[\dots]$ reads
\begin{eqnarray}
\rho_0(x) = \frac{1}{z[\rho_1,\dots,\rho_k]}\int\!\prod_{i=1}^k \!d\rho_i(x_i)\,\,
w(x_1,\dots,x_k)^{m_2}\,\, \delta(x-f(x_1,\dots,x_k))\, ,
\end{eqnarray}
and $z[\rho_1,\dots,\rho_k]$ is fixed by the normalization condition
$\int\!dx\, \rho_{\cdot}(x) = 1$.  The functions $w(x_1,\dots,x_k)$
and $f(x_1,\dots,x_k)$ are model-dependent.  Their explicit form for
the model (\ref{Hamiltonian}) are reported in
App.~\ref{ExplicitFormulae}.

An 1RSB solution is recovered if $Q[\rho]$ is supported on
$\delta$-functions.  We want to compute the stability of this subspace
under the recursion (\ref{2RSBSaddlePoint}). We consider therefore an
order parameter $Q[\rho]$ which is concentrated on narrow
distributions $\rho(x)$. Such distributions can be characterized by
their first two moments:
\begin{eqnarray}
\ox\equiv \int\!d\rho(x)\, x\; ,\; \;\;\; \epsilon \equiv
\int\!d\rho(x)\, x^2 - \left(\int\!d\rho(x)\, x\right)^2\, .
\end{eqnarray}
In the $\epsilon\to 0$ limit we can derive from
Eq.~(\ref{2RSBSaddlePoint}) a recursion for the probability
distribution $Q(\overline{x},\epsilon)$ of these two parameters:
\begin{eqnarray}
Q(\ox,\epsilon) \approx \frac{1}{\hat{\cal Z}}\int\!\prod_{i=1}^{k}\!
dQ(\ox_i,\epsilon_i) \,\, w(\ox_1,\dots,\ox_k)^{m_1}\,\,
\delta\Big(\ox-f(\ox_1,\dots,\ox_k)\Big)\,\,
\delta\Big(\epsilon-\sum_{i=1}^k
\left(\partial_{\ox_i}f\right)^2\epsilon_i\Big)\, .
\label{Linearized}
\end{eqnarray}

Notice that the largest breaking parameter $m_2$ plays no role in the
relation (\ref{Linearized}). Hereafter we shall drop the subscript in
the remaining parameter and set $m=m_1$.  Let us call $\rho_*(\ox)$
the marginal distribution of $\ox$: $\rho_*(\ox)\equiv \int\!
d\epsilon\, Q(\ox,\epsilon)$. It is straightforward to see that this
distribution satisfies the 1RSB equation:
\begin{eqnarray}
\rho_*(\ox) = \frac{1}{z_*}\int\!\prod_{i=1}^k \!d\rho_*(\ox_i)\,\,
w(\ox_1,\dots,\ox_k)^{m}\,\,
\delta\Big(\ox-f(\ox_1,\dots,\ox_k)\Big)\, .
\end{eqnarray}
Therefore $\rho_*(\ox)$ is nothing but the 1RSB solution with
parameter $m$.  If $Q(\ox,\epsilon)\to \rho_*(\ox)\delta(\epsilon)$
under the iteration (\ref{Linearized}), the 1RSB space is stable. In
the opposite case, it is unstable towards 2RSB and (presumably) FRSB.

The distribution $Q(\ox,\epsilon)$ can be represented by a population
of couples $\{(\ox_i,\epsilon_i):i=1,\dots,{\cal N}\}$, and the
recursion (\ref{Linearized}) can be approximated by a population
dynamics algorithm following the ideas of
Ref.~\cite{MezardParisiBethe}. The stability of the 1RSB subspace can
be verified by monitoring a suitable norm of the $\epsilon_i$'s, e.g.
\begin{eqnarray}
\| \epsilon\| \equiv \frac{1}{\cal N}\sum_{i=1}^{\cal N} |\epsilon_i|
\end{eqnarray}
and checking whether $\| \epsilon\|\to 0$ or not. An alternative (and
numerically preferable) procedure is the following. After each sweep
in the population dynamics algorithm the $\epsilon_i$'s are
renormalized: $\epsilon_i \leftarrow \epsilon_i/\lambda$. The
parameter $\lambda$ is chosen such that $\| \epsilon\| = 1$ is kept
fixed (this can be done because Eq.~(\ref{Linearized}) is linear in
the $\epsilon_i$'s). We keep track of $\lambda$ and (eventually)
average it over many iterations. The 1RSB subspace is stable if
$|\lambda|<1$.

\begin{table}
\centering\begin{tabular}{|c|l|}
\hline
$c=4$ & \begin{tabular}{l}
$e_{\rm s}(T) = -1.21771 + 0.06409\, e^{-2\beta}+0.5787\, e^{-4\beta}$\\
$e_{\rm d}(T) = -1.15267 - 0.05816\, T +0.08068\, T^2+
(-0.59409+0.65230\, T)\, e^{-2\beta}$\\
$e_{\rm G}(T) = -1.16667 - 0.04109\, T + 0.10430\, T^2+(-0.63535 + 0.47338\, T)
\,e^{-2\beta}$\\
$e_{\rm ic}(T) = -1.17565 + 0.14471\, e^{-2\beta} - 0.85310\, e^{-4\beta}+
33.618\, e^{-6\beta}$
\end{tabular}\\
\hline
$c=5$ & \begin{tabular}{l}
$e_{\rm s}(T) = -1.39492+0.0994375\, e^{-2\beta} +0.842632\, e^{-4\beta}$\\
$e_{\rm d}(T) = -1.32540-0.03884\, T +(1.79894-3.17138\, T)e^{-2\beta}+
13.2849\, e^{-4\beta}$\\
$e_{\rm G}(T) = -1.35728+0.011316\, T+0.05804\, T^2+
(-0.43749+0.29550\, T)\, e^{-2\beta}$\\
$e_{\rm ic}(T) = -1.35122+0.13262\, e^{-2\beta}+1.4556\, e^{-4\beta}$
\end{tabular}\\
\hline
$c=8$ & \begin{tabular}{l}
$e_{\rm s}(T) = -1.80920+0.07186\, e^{-2\beta} + 0.10056\, e^{-4\beta}+
2.9090\, e^{-6\beta}$\\
$e_{\rm d}(T) = -1.70826-0.04277\, T + (1.0528+0.3575\, T)\, e^{-4\beta}$\\
$e_{\rm G}(T) = -1.77828 - 0.00159\, T +0.07624\, T^2 
+ (-2.7278 + 1.3729\, T)\, e^{-4\beta}$\\
$e_{\rm ic}(T) = -1.76036 + (1.0636 + 0.2779\, T)\, e^{-4\beta}$\\
\end{tabular}\\
\hline
\end{tabular}
\caption{The static ($e_{\rm s}$), dynamic ($e_{\rm d}$), marginal
($e_{\rm G}$) and iso-complexity ($e_{\rm ic}$) energies as obtained
through a population dynamics solution of the 1RSB equations. We
present here a compact representation of these curves as polynomials
in $T$ and $e^{-2\beta}$. The uncertainty due to statistical
fluctuations of the population dynamics algorithm is $\Delta e\lesssim
0.0005$ for $c=4,5$ (and  $\Delta e\lesssim 0.001$ for $c=8$). 
For $e_{\rm ic}(T)$ the systematic error due to the
uncertainty on $\Sigma_{\rm d}(T_{\rm d})$ has to be added.}
\label{CurvesTable}
\end{table}
In Tab.~\ref{CurvesTable} we report the value $e_{\rm G}(T)$ of the
internal energy on the marginally stable states ($|\lambda|=1$) for the
model (\ref{Hamiltonian}) with $p=3$ and several values of $l$.  As
for the the dynamic and static energies, $e_{\rm d}(T)$ and
$e_{\rm s}(T)$, we used populations of size ${\cal N} = 10^5\div 10^6$
and averaged over $200\div 400$ iterations.

One further entry in Tab.~\ref{CurvesTable} is the iso-complexity
energy $e_{\rm ic}(T)$. The corresponding parameter $m_{\rm ic}(T)$ is
defined by the condition
\begin{eqnarray}
\Sigma_T(m_{\rm ic}(T)) = \Sigma_{T_{\rm d}}(m_d(T_d))\, ,
\end{eqnarray}
where, with an abuse of notation, we used the symbol $\Sigma_T(\cdot)$
to denote the complexity as a function of the breaking parameter.  The
relevance of this curve to the issues introduced in
Sec.~\ref{Introduction} was suggested in Ref. \cite{Lopatin} and will
be discussed in the next Section.
%
%
\section{Numerical results}
\label{NumericalSection}

We simulated the model (\ref{Hamiltonian}) for $p=3$ and $c=4,5,8$.
In order to investigate the question in Eq.~(\ref{Question}), we
measured the internal energy of the system, ${\cal O} = {\cal
H}(\sigma)$, both during the relaxation after a quench and during a
slow cooling.

The system is always prepared in a random configuration
($T\!=\!\infty$).  In the quenching experiments the system is let
evolve directly at the final temperature $T_{\rm f}$.  In the
cooling experiments we used the following schedule for the
temperature: $n_{\rm cool}$ Monte Carlo sweeps (MCS) are performed at
each temperature $T \in \{ T_{\rm max}, T_{\rm max}\!-\!\Delta T,
T_{\rm max}\!-\!2\Delta T, \ldots, 2\Delta T, \Delta T, 0\}$.  Values
for $n_{\rm cool}$ are $10^2$, $10^3$, $10^4$ and $10^5$, while those
for $T_{\rm max}$ and $\Delta T$ are reported in the two central
columns of Tab.~\ref{PhaseTransitionTable}.

In order to minimize finite size effects, we considered samples of
size $N=(4\times 10^5-1)$ for quenching experiments, 
and $N=(10^5-1)$ for cooling experiments.  
We checked our results simulating few samples of size $N=(10^6-1)$.
The number of samples varies from $N_{\rm s}=5$ for
quenching experiments to $N_{\rm s}\gtrsim 10$ for cooling experiments.

\begin{figure}
\begin{center}
\includegraphics[width=0.75\textwidth]{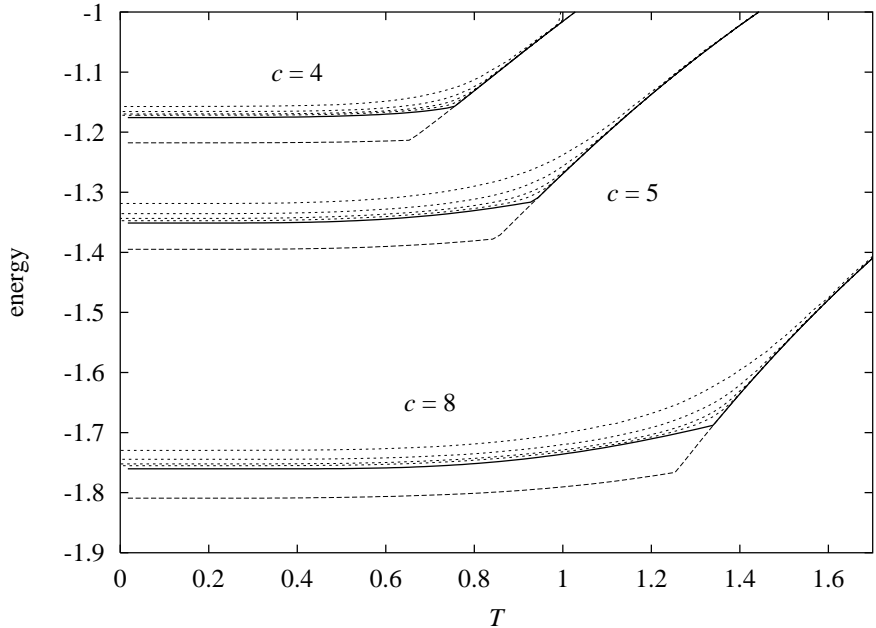}
\end{center}
\caption{Cooling experiments for several values of the connectivity
$c$ and of the cooling rate $n_{\rm cool}$. Dotted lines represent the
numerical results for $e_{\rm cool} (T,n_{\rm cool})$.  Dashed lines
correspond to the analytical result for the equilibrium energy, and
solid lines to the paramagnetic energy (for $T>T_{\rm d}$) or to the
iso-complexity energy $e_{\rm ic}(T)$ (for $T<T_{\rm d}$).}
\label{cool_all}
\end{figure}

In Fig.~\ref{cool_all} we show the internal energy of the system as a
function of the temperature during the cooling experiments with 4
different values of $n_{\rm cool}$ for each connectivity (upper dotted
lines).  It should be clear from this picture that the system is
undergoing a dynamic arrest preventing it to reach the
thermodynamic energy (lower dashed line).  We also show in this
picture the iso-complexity energy (bold solid line).  Numerical
evidences, to be shown below, strongly suggests that the system
follows the iso-complexity line when cooled infinitely slowly.

\begin{figure}
\begin{center}
\includegraphics[width=0.75\textwidth]{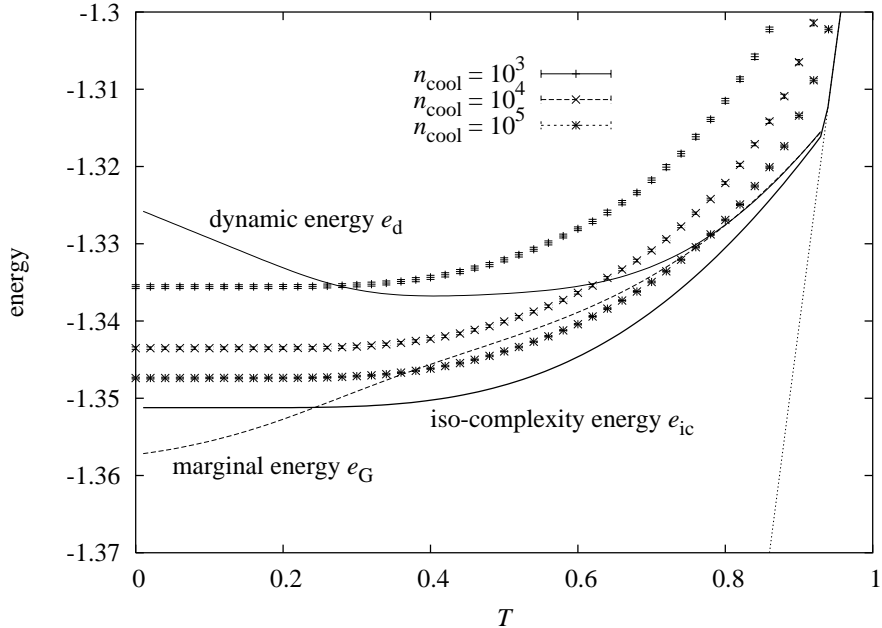}
\end{center}
\caption{A closer look at the cooling experiments for connectivity
$c=5$.}
\label{cool_c5_zoom}
\end{figure}

In order to better study the $n_{\rm cool} \to \infty$ limit we choose
one connectivity, e.g.\ $c=5$, and we zoom on the interesting region.
Figure \ref{cool_c5_zoom} shows cooling energies for $c=5$ on a
different scale ($n_{\rm cool}=10^2$ has been omitted for clarity).
Statistical errors are smaller than symbol size.  For comparison we also plot
(from top to bottom) the analytic curves corresponding to the
threshold or dynamic energy, $e_{\rm d}(T)$, the marginal energy,
$e_{\rm G}(T)$, and the iso-complexity energy, $e_{\rm ic}(T)$, reported
in Tab.~\ref{CurvesTable}.

It is expected that energy relaxation converges to a threshold energy, 
lying between $e_{\rm G}(T)$ and $e_{\rm
d}(T)$.  This belief is based on the fact that for energies below
$e_{\rm G}(T)$ 1RSB states are stable.  On the contrary we clearly see
from Fig.~\ref{cool_c5_zoom} that a cooling experiments may bring the
system to an energy below $e_{\rm G}(T)$.  In Sec.~\ref{DiscussionSection} 
we will argue that this numerical observation is not in contrast with the
existence of well defined 1RSB states.

Next we ask whether for very
slow cooling rates, i.e.\ for $n_{\rm cool} \to \infty$, the energy of
the system follows the iso-complexity energy.  In order to investigate this
point we extrapolate the cooling energy for $n_{\rm cool} \to
\infty$ as follows. For any temperature $T$ we fit the data
to the law
\[
e_{\rm cool}(T,n_{\rm cool}) = e_{\rm cool}(T) + a\; n_{\rm
cool}^{-b}\;.
\]
For all the connectivities considered ($c=4,5,8$) and for 
temperatures $T \alt T_{\rm d}$ typical values for the best fitting $b$
parameter are in the range $0.3 \div 0.36$.

\begin{figure}
\begin{center}
\includegraphics[width=0.75\textwidth]{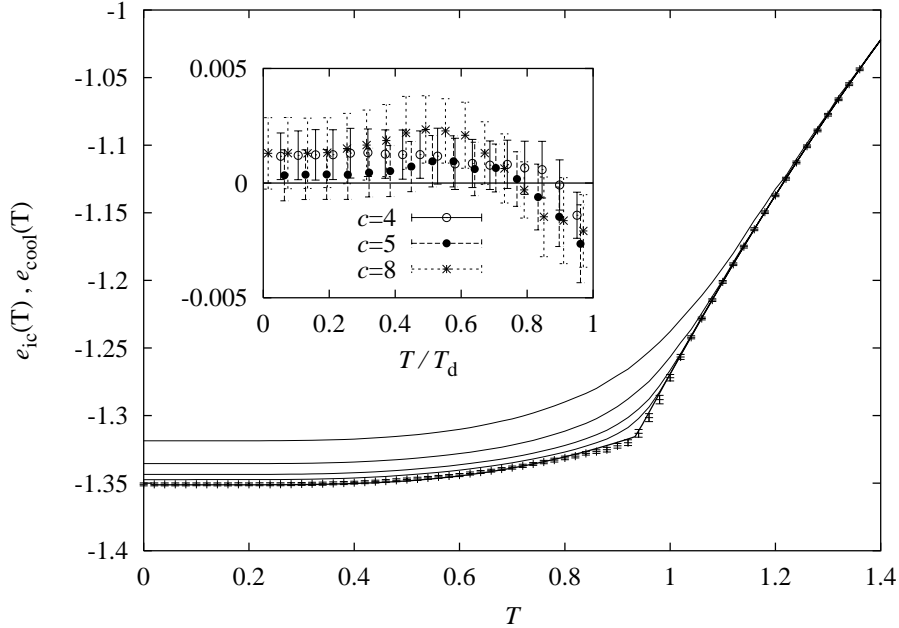}
\end{center}
\caption{Extrapolation of the cooling experiments to infinitely slow
cooling rate. The crosses with error bars in the main frame represent
the extrapolated data. Error bars include the uncertainty in the
analytical calculation of $e_{\rm ic}(T)$. In the inset: difference
between the extrapolated data and the analytical result for $e_{\rm
ic}(T)$, for connectivities $c=4,5,8$.}
\label{extrap_c5}
\end{figure}

The extrapolated energies $e_{\rm cool}(T)$ for $c=5$ are plotted in
the main panel of Fig.~\ref{extrap_c5} (points with errors).  We have
included in the error bars also the contribution due to the
uncertainty on the analytic estimation of $e_{\rm ic}(T)$.  Thin lines
are cooling energies with $n_{\rm cool}=10^2,10^3,10^4,10^5$, while
the thick line is the iso-complexity energy.

Within the statistical error $e_{\rm cool}(T)$ and $e_{\rm ic}(T)$
perfectly coincide, as can been better seen in the inset of
Fig.~\ref{extrap_c5}, where the difference $e_{\rm cool}(T)-e_{\rm
ic}(T)$ has been plotted for all the connectivities.

Let us now consider the energy relaxation after a sudden quench to
temperature $T$.  The temperatures studied for each connectivity are
written in the last column of Tab.~\ref{PhaseTransitionTable}. Each
simulations has been run for $10^6$ MCS.  We assume that on late times
the energy relaxation may be approximated by a single power law
behavior
\be
e_{\rm quench}(T,t) = e_{\rm quench}(T) + a'\; t^{-b'}\;.
\label{fit}
\ee
Best fitting values for $b'$ are typically around $1/3$, for all the
connectivities and the temperatures studied.

The extrapolation of $e_{\rm quench}(T)$ depends of course on the
range of times used in the fitting procedure, because of sub-leading
corrections.  In order to have an appreciation of this effect, we
fitted the $e_{\rm quench}(T,t)$ data to expression (\ref{fit}) within
the interval $t \in [t_{\rm min},t_{\rm max}]$, with $t_{\rm max} \ge
10^5$ and $t_{\rm max} / t_{\rm min} \ge 200$.

\begin{figure}
\begin{center}
\includegraphics[width=0.75\textwidth]{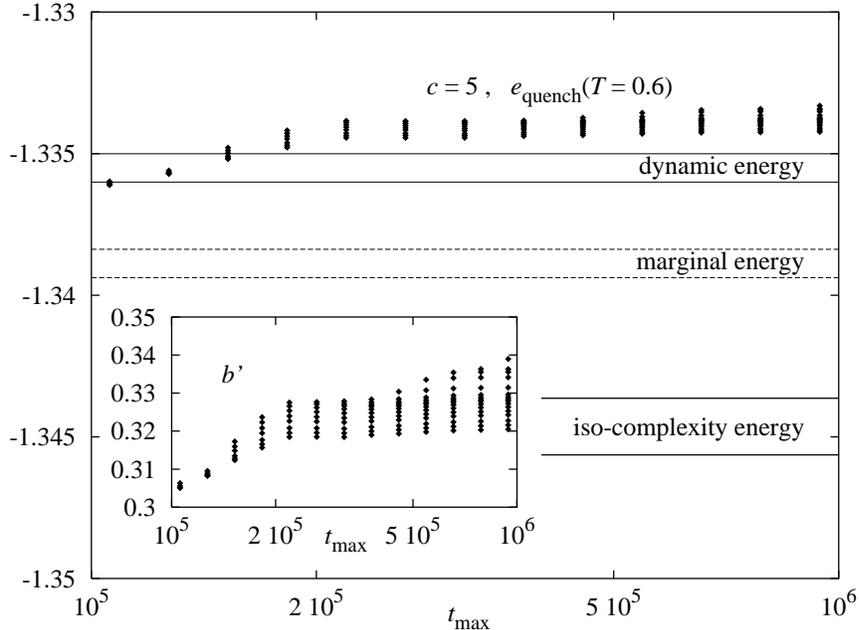}
\end{center}
\caption{The energy extrapolated from the quenches (main panel) and
the best fitting $b'$ parameter (inset), for $c=5$ and $T=0.6$.  Each
analytical energy is shown with 2 horizontal lines corresponding to
its mean plus or minus one standard deviation.}
\label{ex_relax_c5}
\end{figure}

In Fig. ~\ref{ex_relax_c5} the extrapolated energy $e_{\rm quench}(T)$
(in the main panel) and the best fitting $b'$ (in the inset) are shown
as a function of $t_{\rm max}$ for $c=5$ and $T=0.6$ (very similar
results has been obtained for the remaining connectivities and
temperatures).  For each value of $t_{\rm max}$, different points
correspond to different values of $t_{\rm min}$.  Statistical errors
(not shown for clarity) are of the same order of the spread of points.
Results seems to be almost $t_{\rm min}$-independent, and also the
dependence on $t_{\rm max}$ is rather mild.

From Fig.~\ref{ex_relax_c5} we conclude that the extrapolated energy
value is reasonably robust.  Moreover the comparison with analytic energies
supports the conclusion that $e_{\rm quench}(T)$ stays above the energy
$e_{\rm ic}(T)$ reached by cooling experiments.

On the other hand we expect $e_{\rm quench}(T)$ to coincide with the
FRSB dynamic energy $e_{\rm d,FRSB}(T)$, i.e.\ with the maximal
energy such that the complexity is strictly positive, once FRSB
effects have been taken into account~\cite{NostroInst}.  While we know
that $e_{\rm G}(T) \le e_{\rm d,FRSB}(T) \le e_{\rm d}(T)$, we do not
yet have any estimate of this threshold. Verifying this theoretical
expectation is therefore very difficult. Based on the data of
Fig.~\ref{ex_relax_c5} we guess that, if our expectation proves
correct, then $e_{\rm d,FRSB}(T)$ is `close' to $e_{\rm d}(T)$ in the
models at hand.  The vicinity of the energy extrapolated from the
quenches $e_{\rm quench}(T)$ to the dynamic energy $e_{\rm d}(T)$ is
well verified for all the connectivities and the temperatures studied
in this work (see last column of Tab.~\ref{PhaseTransitionTable}).
This is presumably a general property of the model (\ref{Hamiltonian})
for the range of connectivities considered in this paper.

%
%
\section{Discussion}
\label{DiscussionSection}

We think that previous pages provide solid support for three
statements concerning the general question (\ref{Question}): $(i)$
$e_{\rm cool}(T)<e_{\rm quench}(T)$; $(ii)$ $e_{\rm quench}(T) >
e_{\rm G}(T)$; $(iii)$ $e_{\rm cool}(T) = e_{\rm ic}(T)$.  For certain
values of the temperature also a fourth statement holds: $(iv)$
$e_{\rm cool}(T) < e_{\rm G}(T)$.  The first of these statements, in
particular, implies a negative answer to the question
(\ref{Question}).  While the second statement was largely
expected~\cite{NostroInst}, the fourth one comes as a surprise, being
in contradiction with some widespread expectation in the field.
Notice moreover that these statements gets sharper as the temperature
is lowered. In fact the separation between $e_{\rm d}(T)$, 
$e_{\rm ic}(T)$ and  $e_{\rm G}(T)$ increases at low temperature, 
cf.\ Fig.~\ref{cool_c5_zoom}.

As for the third one, namely $e_{\rm cool}(T) = e_{\rm ic}(T)$,
there exists a simple argument which helps to understand this
relation. Consider a system with 1RSB, and assume that no temperature
chaos is present.  In other words, pure states can be traced in
temperature without crossings or bifurcations. It is then easy to show
\cite{Lopatin} that the energy of a pure state, regarded as a function
of the temperature, follows an iso-complexity line (the value of the
complexity on this line depending upon the state). In particular,
$e_{\rm ic}(T)$ is the energy of the first metastable states to appear
when the temperature is lowered across $T_{\rm d}$. Therefore, as far
as single-time quantities are concerned, the system behave {\it as if\ }
it equilibrated within the first metastable states encountered in its
cooling history.  Of course this picture is too simplified, and in
fact the system ages also if it is slowly cooled across $T_{\rm d}$
\cite{AgingSimul}. A pictorial view of the evolution of
the system during a slow cooling given in Fig.~\ref{Pictorial}.

Coming back to the discussion of metastable states 
in Sec.~\ref{Introduction}, the present analysis suggest a general 
(and testable) strategy for tailoring out-of-equilibrium
ensembles of the form (\ref{Constrained}). The general idea is to add
external sources $\lambda_1$, $\lambda_2$, etc conjugated
to extensive observables ${\cal O}_1$, ${\cal O}_2$, etc.
Such observables could be, for instance, the energy or the magnetization.
The important rule is that any slow change in the control parameters
induces an iso-complexity change in the state of the system. 
This is analogous to entropy conservation for adiabatic
transformations in classical thermodynamics. Although 
here the system is in contact with a thermal bath, slow degrees
of freedom are effectively thermally isolated.

Let us finally comment on the relevance of our results for the
analysis of local search algorithms for random combinatorial
optimization problems (here we make the usual identification between
cost function and Hamiltonian).  It has become customary
\cite{PTAC,XorSat,DynamicCodes} to compare the asymptotic cost $e_{\rm
search}(t\to\infty)$ achieved by such algorithms with the 1RSB
threshold energy $e_{\rm d}(T=0)$ . Here (see also \cite{PTAC} for a
discussion concerning this point) we notice that:
\begin{enumerate}
\item In general, the local search algorithm does not satisfy detailed
balance with respect to the Boltzmann distribution at any temperature
$T$.  If this is the case, the very existence of a relation between
$e_{\rm search}(t\to\infty)$ and any thermodynamic quantity [and {\it
a fortiori} $e_{\rm d}(T=0)$] is an open problem.
\item In the simplest case, the local search algorithm satisfies
detailed balance at any time $t$ with respect to the Boltzmann
distribution at temperature $T(t)$, with $T(t)$ a deterministic
schedule such that $T(t)\to 0$ as $t\to\infty$. Surprisingly, even in
this case $e_{\rm d}(T=0)$ is not relevant. However, our results seem
to imply that $e_{\rm search}(t\to\infty)\ge e_{\rm ic}(T=0)$.  This
is an encouraging remark for the application of such algorithms.
First of all $e_{\rm ic}(0)<e_{\rm d}(0)$. Moreover one could imagine
constructing more complex annealing paths using clever deformations
of the cost function to define the Hamiltonian.  A smart deformation
would probably allow to reduce $e_{\rm ic}(0)$.
\end{enumerate}
%
%
\acknowledgments

We are grateful to Andrei Lopatin for bringing to our attention the
relevance of the iso-complexity curve, and to Giorgio Parisi for
stressing the importance of assuming the absence of temperature chaos.
This work was partially supported by the ESF programme SPHINX and by
the European Community's Human Potential Programme under contracts
HPRN-CT-2002-00307, Dyglagemem, and HPRN-CT-2002-00319, Stipco.

%
%
\appendix
\section{Some formulae for the diluted $p$-spin model}
\label{ExplicitFormulae}

For the sake of self-containedness we report here some explicit
formulae for the diluted $p$-spin model (\ref{Hamiltonian}). These
formulae can be used in computing the 1RSB stability threshold along
the lines of Sec.~\ref{ModelSection}.

It is convenient to express the cavity equations in terms of two
functional order parameter $Q[\rho]$ and $\Qh[\rh]$. The first one is
related to the distribution of cavity fields when one {\it
interaction} term is removed from the system. The second one
corresponds to the distribution of cavity fields when one {\it spin}
is removed from the system.

The 2RSB cavity equations read
\begin{eqnarray}
Q[\rho] & =  & \frac{1}{\cal Z}\int\!\prod_{i=1}^{l}\! d\Qh[\rh_i] \,\, 
z[\rh_1,\dots,\rh_l]^{m_1/m_2}\,\, \delta\left[\rho-\rho_0[\rh_1,\dots,\rh_l]\right]\, ,\\
\Qh[\rh] & =  & \int\!\prod_{j=1}^{p-1}\! dQ[\rho_j] \,\, {\mathbb E}_J
\,\, \delta\left[\rh-\rh_0[J;\rho_1,\dots,\rho_{p-1}]\right]\, ,
\end{eqnarray}
where ${\mathbb E}_J$ denotes the expectation with respect to the random 
variable $J$ which takes values $+1$ or $-1$ with equal probability).
The mappings $\rho_0[\dots]$ and $\rh_0[\dots]$ are defined below:
\begin{eqnarray}
\rho_0(x) & =&\frac{1}{z[\{\rh_i\}]}\int\!\prod_{i=1}^{l}\! d\rh_i(y_i)
\; w(\{y_i\})^{m_2}\;\delta(x-y_1-\dots-y_l)\, ,\\
\rh_0(y) & = & \int\!\prod_{j=1}^{p-1}\! d\rho_j(x_j)\;
\delta\left(y-\frac{1}{\beta}\, {\rm atanh}[\tanh\beta J\tanh\beta x_1\dots
\tanh\beta x_{p-1}]\right)\, .
\end{eqnarray}
The reweighting factor is
\begin{eqnarray}
w(y_1,\dots,y_l) = \frac{2\cosh(\beta\sum_{i=1}^l y_i)}
{\prod_{i=1}^l2\cosh\beta y_i}\, .
\end{eqnarray}
%
%
\section{The large connectivity limit}
\label{LargeConnectivity}

The recipe we proposed for computing the stability condition of a
diluted spin glass model does not necessarily capture the most
relevant instability. In this Appendix we show that, in the large
connectivity limit $l\to\infty$, our approach yields the replicon
instability already computed in \cite{Gardner}. This provides an
important check of our calculation.

When adapting the basic recursion (\ref{Linearized}) to the cavity
equations reported in the previous Appendix, it is necessary to use
{\it two} distributions $Q(\ox,\epsilon)$ and $\Qh(\oy,\eh)$.  These
distributions can be used to define the functions $\epsilon(x)$ and
$\eh(y)$ as follows
\begin{eqnarray}
\epsilon(x) = \int\! d\epsilon'\; Q(x,\epsilon')\; \epsilon'\, ,\;\;\;\;\;
\;\;\;\;\;\eh(y) = \int\! d\eh'\; \Qh(y,\eh')\; \eh'\, .
\end{eqnarray}
It is easy to show that Eq.~(\ref{Linearized}) implies a recursion of
the type $[\epsilon(x),\eh(y)]\mapsto [\epsilon'(x),\eh'(y)]$, where:
\begin{eqnarray}
\epsilon'(x)  =   \int\! dy\;\; M(x,y)\; \eh(y)\, ,\;\;\;\;\;\;\;\;
\eh'(y)  =   \int\! dx\;\; \Mh(y,x)\; \epsilon(y)\, .
\label{Recursion}
\end{eqnarray}
The kernels of this mapping are given in terms of the 1RSB solution 
$\rho_*(x)$, $\rh_*(y)$:
\begin{eqnarray}
\hspace{-0.75cm}
M(x,y) & = & \frac{l}{z_*}\, \int\!\prod_{i=1}^{l-1}\!d\rh_*(y_i)\;
w(y,y_1\dots y_{l-1})^m\; \; \delta(x-y-y_1-\dots - y_{l-2})\, ,
\label{Kernel1}\\
\hspace{-0.75cm}
\Mh(y,x) & = &  \int\!\prod_{j=1}^{p-2}\!d\rho_*(x_j)\;
\Delta(x,\{x_j\})\; \delta\left(y-\frac{1}{\beta}\, {\rm atanh}
[\tanh\beta J\tanh\beta x \cdots\tanh\beta x_{p-2}]\right)\, ,
\label{Kernel2}
\end{eqnarray}
where $z_* = z[\rh_*\dots\rh_*]$ and
\begin{eqnarray}
\Delta(x,\{x_j\}) \equiv (p-1)\, \left[\frac{\tanh\beta
J(1-\tanh^2\beta x)\prod_{j=1}^{p-2}\tanh\beta x_j} {1-\tanh^2\beta
J\tanh^2\beta x\prod_{j=1}^{p-2}\tanh^2\beta x_j}\right]^2\, .
\label{Kernel2Reweight}
\end{eqnarray}
Notice that we do not need to average on the sign of $J$ in
Eqs.~(\ref{Kernel2}) and (\ref{Kernel2Reweight}) because the
distribution $\rho_*(x)$ is symmetric.

The stability of the fixed point $\epsilon(x) = 0$, $\eh(y) = 0$ under
the recursion (\ref{Recursion}), can be determined by diagonalizing
the composition of the kernels (\ref{Kernel1}), (\ref{Kernel2}). As we
will see shortly, this diagonalization becomes considerably easier in
the limit $l\to\infty$.

First of all, we rescale the coupling strength by setting $J =
\sqrt{p/2l}$ in such a way that the energy remains finite in the large
connectivity limit. We also rescale the cavity fields defining $y =
\sqrt{p/2l}\; \tilde{y}$ (hereafter we shall drop the tilde).  It is
easy to see that, in the large connectivity limit, the 1RSB fields
distributions become
\begin{eqnarray}
\rho_*(x) & = & \frac{1}{\cal Z}\; 
(\cosh\beta x)^m \;\; e^{-x^2/2\lambda}\, ,\label{LimitRho}\\
\rh_*(y) & = & \int\!\prod_{j=1}^{p-1}\!d\rho_*(x_j)\;\;
\delta(y-\tanh\beta x_1\cdots\tanh\beta x_{p-1})\, . 
\end{eqnarray}
The parameter $\lambda$ must be found by solving the equation
\begin{eqnarray}
\lambda = \frac{p}{2}\, \left[\int\!d\rho_*(x)\;\tanh^2\beta
x\right]^{p-1}\, .
\end{eqnarray}
This equation was already found in Ref. \cite{Gardner} while solving
the fully connected $p$-spin model.

Taking the $l\to\infty$ limit also in the kernels (\ref{Kernel1}) and
(\ref{Kernel2}) we obtain
\begin{eqnarray}
M(x,y) & = & \frac{p}{2}\; \rho_*(x)\, ,\label{LimitKernel1}\\
\Mh(y,x) & = & \int \!\prod_{j=1}^{p-2}\!d\rho_*(x_j)\;
D(x,\{x_j\})\; \delta(y-\tanh\beta x \cdots\tanh\beta x_{p-2})\, ,
\end{eqnarray}
with
\begin{eqnarray}
D(x,\{x_j\}) = \left[\beta(1-\tanh^2\beta
x)\prod_{i=1}^{p-2}\tanh\beta x_i \right]^2\, .
\end{eqnarray}
It is evident from Eq.~(\ref{LimitKernel1}) that there is a unique
non-vanishing eigenvalue, and that the corresponding eigenvector has
$\epsilon(x) \propto\rho_*(x)$. A little thought shows that the
eigenvalue is
\begin{eqnarray}
\Lambda = \frac{1}{2}\, p(p-1)\beta^2 q^{p-2}\int\! d\rho_*(x) 
\; (\cosh\beta x)^{-4}\, ,
\end{eqnarray}
with $\rho_*(x)$ given by Eq.~(\ref{LimitRho}). As anticipated, this
result coincides with the replicon eigenvalue of \cite{Gardner}.
%
%

\end{document}